\documentclass[aps,prb,twocolumn,reprint,superscriptaddress,showkeys,floatfix,amsmath,amssymb]{revtex4-1}
\usepackage{float}
\usepackage[latin9]{inputenc}
\usepackage{color}
\usepackage{booktabs}

\usepackage{graphicx}

\usepackage{natbib}
\usepackage{csquotes}
\usepackage{xcolor, soul}
\usepackage{dcolumn}
\usepackage{bm}
\newcommand{\rtcom}[1]{\hl{}}{}

\usepackage{float}

\usepackage{float}
\begin{document}
	
	\preprint{APS/123-QED}
	
	\title{ Interfacial effect on the optoelectronic and piezoelectric properties of Ge-Sn terminated Halide Perovskite heterostructure from first-principles study}
	
	\author{L. Celestine}
	\affiliation{Advanced Functional Materials \& Simulation Lab (AFMSL), Department of Physics, Aizawl-796004, India}
	\affiliation{Physical Sciences Research Center (PSRC), Department of Physics, Pachhunga University College,  Aizawl-796001, India}
	\author{R. Zosiamliana}
	\affiliation{Advanced Functional Materials \& Simulation Lab (AFMSL), Department of Physics, Aizawl-796004, India}
	\affiliation{Physical Sciences Research Center (PSRC), Department of Physics, Pachhunga University College,  Aizawl-796001, India}
	\author{H. Laltlanmawii}
	\affiliation{Advanced Functional Materials \& Simulation Lab (AFMSL), Department of Physics, Aizawl-796004, India}
	\affiliation{Physical Sciences Research Center (PSRC), Department of Physics, Pachhunga University College,  Aizawl-796001, India}
	\author{B. Chettri}
	\affiliation{Advanced Functional Materials \& Simulation Lab (AFMSL), Department of Physics, Aizawl-796004, India}
	\author{Lalhum Hima}
	\affiliation{Physical Sciences Research Center (PSRC), Department of Physics, Pachhunga University College,  Aizawl-796001, India}
	\author{Lalhriat Zuala}
	\affiliation{Physical Sciences Research Center (PSRC), Department of Physics, Pachhunga University College,  Aizawl-796001, India}
	\author{S. Gurung}
	\affiliation{Physical Sciences Research Center (PSRC), Department of Physics, Pachhunga University College,  Aizawl-796001, India}
	\author{A. Laref}
	\affiliation{Department of Physics and Astronomy, College of Science, King Saud University, Riyadh, 11451, Saudi Arabia}%
	\author{D. P. Rai}
	\email[D. P. Rai:]{dibyaprakashrai@gmail.com}
	\affiliation{Advanced Functional Materials \& Simulation Lab (AFMSL), Department of Physics, Aizawl-796004, India}
	
	\begin{abstract}
		Since the very early stages of research on sustainable technologies, green energy conversion has always been a prime focus. With the discoveries of countless functional materials in recent years, significant progress has been made to meet the global energy demand for sustainable development. Among them, halide perovskites have emerged as one of the most promising and reliable materials. In this work, we have investigated the lead-free halide perovskites $vis$ CsGeCl$_3$ and RbSnBr$_3$ within a framework of density functional theory (DFT) to explore their potential applicability in harvesting clean and renewable energy. This study gives a comprehensive analysis of the bulk, surface (001), and Ge-Sn-terminated interfaces within GGA and mGGA functionals. Interestingly, the inherent asymmetric arrangements of the systems exhibit remarkable optoelectronic and piezoelectric properties. The piezoelectric performance of each surface cut has been validated through the electromechanical coupling calculation. 
	\end{abstract}
	
	\maketitle
	
	\section{Introduction}
	Overpopulation and numerous ambitious developmental projects are the key routes to excessive energy consumption that ultimately lead to a surge in the global energy demand \cite{Ang2022}. The rapid depletion of conventional energy resources like fossil fuels raised concerns about the sustainability. Foreseeing the potential threats to life sustainability and the looming ecosystem has urged the government and other stakeholders have urged the development of green energy-generation and high energy-capacity storage systems for future use\cite{Qazi2019}. To solve this energy crisis, the current research is extensively oriented towards the development of an energy harvesting system via materials modeling. The discoveries of halide perovskites (HPs) have emerged as a new hope for materials science and materials engineering, as they have shown promise to contribute to the advancement of new technology, such as in solar cells\cite{Wang2021}, light-emitting diodes (LEDs)\cite{Liu2020}, gas sensors\cite{Paul2023}, memory devices\cite{Satapathi2022}, etc. Attention has been drawn to utilising halide-based perovskites, and researchers have delved into theoretical and experimental approaches for various purposes using HPs. Within the family, the ones containing lead(Pb) element, such as CsPbCl$_3$\cite{Nabi2021}, CsPbBr$_3$\cite{He2021}, CsPbI$_3$\cite{Kaplan2023}, RbPbCl$_3$\cite{Huang2018}, RbPbBr$_3$\cite{Ghosh2024}, RbPbI$_3$\cite{Yu2024} possess great stability with higher responses in the field of optoelectronics, thermoelectric, and piezoelectric applications. But these are reported to be harmful to the environment. So, instead of utilizing hazardous material, replacement of Pb-element with Germanium (Ge) or Tin (Sn) worked out as well \cite{Celestine2025}. Pb-free HPs tremendously gained interest among scientists and industries to produce items which are eco-friendly \cite{Celestine2024}.
	
	\par Generally, halide perovskites have become promising candidates to be used in high-performance absorber layers in solar cells due to their exceptional capacity to absorb the electromagnetic radiation (Sun) and generate electron-hole pairs\cite{Jung2017, Bhattarai2023}. Also, as a hole transport layer (HTL)\cite{Xu2023, J-YLim2022} due to their capabilities to enhance the material's stability, block electrons, and promote the hole movement. Surface studies towards HPs have been carried out to enhance the material's properties, mostly for solar cell applications. Haruyama et.al\cite{Haruyama2016} have pursued their interest towards organic-based halide perovskite and reported CH$_3$NH$_3$PbI$_3$(MAPbI$_3$) with (001) and (110) surfaces as their most probable electron transport material to improve the performance of perovskite solar cell (PSC). Giovanni et.al\cite{DiLiberto2021a} studied CsPbX$_3$(001) and (110) by designing an interface of CsPbX$_3$ with TiO$_2$ (CsPbX$_3$/TiO$_2$) for solar energy harvesting. Jose et.al\cite{Laranjeira2023b} investigated CsPbX$_3$ (001) by designing and examining the influence of anion hardness for various applications. Furthermore, Young et.al\cite{Jung2017} predicted the absolute energy levels under the influence of cation exchange (Cs)RbSnI$_3$ (001) surface.
	
	\par Recent studies have shown Pb-free halide-based perovskites are promising multifunctional power harvesting materials. Not only in solar power generation but also through thermoelectric power generators \cite{Hsu2021}, wind power generators by incorporating with Li-batteries\cite{Alan2021}, photorechargeable batteries\cite{Tewari2025}, and so on. Despite capturing high power-energy, these are obtained through the influence of external environmental forces, which in turn give some limitations. To avoid these drawbacks, an alternate method to fulfil the criterion of generating and storing energy is a must \cite{Rajan2018}. A novel piezoelectric technique, which does not require the influential attributes of the environment, was developed by the Curie brothers in 1880\cite{Martin1972}. This phenomenon contributes to the development of converting mechanical stress into green and clean renewable electrical energy with a cost-effective method \cite{Elahi2018}. Generation of energy via the piezoelectric technique using HPs has tremendously increased in recent years. Sun et.al\cite{Sun2025} studied the piezocatalytic characteristics of PbTiO$_3$ (PTO) in bulk and layered-surfaces (001), then found that some layers of PTO piezoresponses are less than the bulk's response, which is caused by the depolarization effect (reverse transfer of electrons). Liu et.al\cite{Liu2018} investigated the performance of the interfacial ferroelectric polarization in CsPbI$_3$/CsSnI$_3$ heterostructure and reported their results to have potential use in optoelectronics purposes. Furthermore, Xue et.al\cite{Xue2022} recorded a high output voltage of self-polarization piezoelectric nanogenerator (PENG) in PVDF/CsPbBr$_3$. 
	
	\par To the best of our knowledge, obtaining a semiconducting heterostructure using non-centrosymmetric inorganic Pb-free HPs for optoelectronic and piezoelectric purposes has been a challenging topic and scarcely reported in both theoretical and experimental studies. Knowing this research gap, this work focuses on having a heterostructure with our two proposed compounds $\sim$ CsGeCl$_3$ and RbSnBr$_3$, and investigate their stabilities and towards energy-related purposes. At first, we investigate the said compounds' bulk properties, which is followed by cleaving the materials. We confirmed that the cleavage plane along the (001) direction is the most probable surface(slab) to obtain for both materials due to the lower surface energy and better stability, as confirmed by our study and literature surveys. Then, we examine these surfaces' stabilities and potential characteristics. Consequently, we form a heterostucture using the two surfaces obtained and proceed with further investigation. Through this novel approach, we thus obtain interesting results which suit various technological aspects.

	\section{Computational Details}
	\par Our first principles investigation of interfacial- CsGeCl$_3$/RbSnBr$_3$  heterostructure study is based on the density functional theory (DFT)\cite{Hohenberg1964} implemented in QuantumATK Q-2019.12\cite{Smidstrup2019}, which relies on the linear combination of atomic orbitals (LCAO) technique\cite{Schlipf2015}. To accurately account for electron-ion interactions, two well-known approximations $\sim$ generalized gradient approximation (GGA)\cite{Perdew1966} and meta-generalized gradient approximation (mGGA)\cite{Sun2011} were adopted. An optimizer algorithm drawn from Newton-quasi technique named the Limited Memory Broyden-Fletcher-Goldfarb-Shanno (LBFGS) has been used to carry out full cell optimization\cite{Zhao2021}. Throughout the cell optimization, we have relaxed the cell's parameters, which include the volumes, space groups and atomic positions. To ensure the structural and cell energy convergence criterion, the stress tolerance and Hellmann-Feynman force were set to 0.00001 eV \AA \ and 0.01 eV \AA \ with a maximum step size of 500, respectively. For each atom (Cs, Rb, Ge, Sn, Cl, and Br), a PseudoDojo potential, which is similar to double zeta polarized (DZP), were adopted. To balance the accuracy and computational cost, a Monkhorst-Pack scheme\cite{Monkhorst1976} of 8$\times$8$\times$8 (8$\times$8$\times$1) k-mesh points was sampled for geometry optimization and with a denser 12$\times$12$\times$12 (12$\times$12$\times$1) k-points for the ground state properties throughout the study. Using the cleavage cut tool, we obtain the surface slab models of CsGeCl$_3$ (001) and RbSnBr$_3$ (001), respectively. Employing the moment tensor potential with Maxwell-Boltzmann distribution at 300 K, molecular dynamics (MD) simulations\cite{Charles1989} have been used to verify the thermal stability of the compounds under investigation. Throughout the simulations, a total number of 5000 steps with 1 femtosecond (fs) time step, equivalent to 0.001 picosecond (ps), were allotted for the bulk systems. Additionally, for the constructed surface slabs and interface systems, we calculate their respective surface energies \textit{(E$_{surface}$)} and interfacial energies ($\gamma$). The QuantumATK code utilises a Berry-phase polarization technique to compute piezoelectric tensors\cite{Bechmann1958}. During the tensors computation, samplings of k-points were set up using a Monkhorst-Pack Grid for self-consistent calculation and integration along a $\sim$ 11 $\times$ 5 $\times$ 5, b $\sim$ 5 $\times$ 11 $\times$ 5, and c$\sim$ 5 $\times$ 5 $\times$ 11, respectively. Further computational details for elastic properties are added in the supplementary file (ESI).
	
	\section{Results and Discussion}
	\subsection{Bulk Properties}
	\subsubsection{Structural \& stability, electronic and optical properties}
	\par In this paper, we investigated our proposed halide-based perovskites in their optimal and lowest possible energy. All the structural information is taken from the online open-access database "Materials Project"\cite{Jain2013}. Our systems, CsGeCl$_3$ and RbSnBr$_3$, follow the well-known perovskite chemical formula of ABX$_3$. Both were studied in different phases CsGeCl$_3$ (Trigonal)\cite{Thiele1987} and RbSnBr$_3$ (Orthorhombic)\cite{Sabine2016}. The optimized cells' parameters and volumes, after fully achieving their minimal energies, are obtained and tabulated in Table \ref{Table 1}, whose values are in good agreement with previously reported data. Detailed structural parameters, such as lattice parameters and volumes, are kept in S1. 
	
	\begin{figure}[hbt!]
		\small
		\includegraphics[height=3cm]{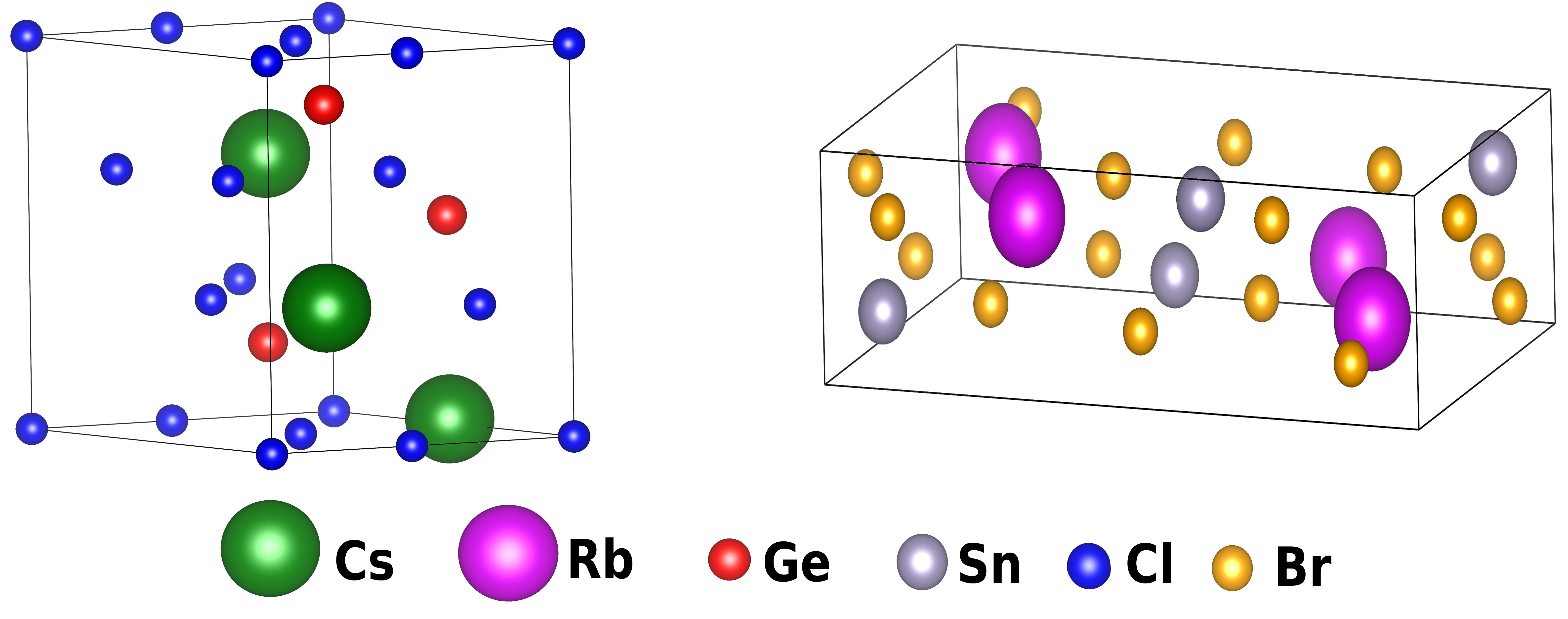}
		\caption{Schematic 3D figures of the bulk perovskites}
		\label{Fig.1Structure}
	\end{figure}

	\par In order to assess the 3D halide perovskites' stability and crystalline formation, we utilised Goldschmidt's tolerance factor "t"\cite{Bartel2019}
	\begin{equation}
		t = \dfrac{r_A + r_B}{\sqrt{2}(r_B + r_X)}
		\label{Eq 1}
	\end{equation}
	
	\par where r$_A$, r$_B$, and r$_X$ being the ionic radii\cite{Shannon1976} of atoms A(Cs/Rb), B(Ge/Sn), and X(Cl/Br). Therefore, the tolerance factors of CsGeCl$_3$ and RbSnBr$_3$ are computed using Eq.\ref{Eq 1} and tabulated in Table \ref{Table 1}. Depending upon the crystalline orientations, the \textit{t}-values can also differ. From Table \ref{Table 1}, we observed that the \textit{t}-value of CsGeCl$_3$ is 1.11( \textgreater1), indicating a hexagonal structure. But for RbSnBr$_3$, the \textit{t}-value is 0.89($\sim$0.71-0.9) which reveals an orthorhombic crystal arrangement. These results are in good agreement with others' work.
	
	\begin{table*}[hbt!]
		\centering
		\caption{\ Calculated values of ionic radii (\textit{r}) in \AA \,  tolerance factor (\textit{t}), and band gaps (E$_g$)  of CsGeCl$_3$ and RbSnBr$_3$} 
		\label{Table 1}
		\begin{tabular*}{\textwidth}{@{\extracolsep{\fill}}l|l|l|l|l|l|l}
			\hline
			\textbf{Atoms} & \textit{r} & \textbf{Compounds} & \textit{t} & Other's value& E$_\textit{g}$&Other's value\\
			\hline
			\textbf{Cs$_{(A_{1})}$ } & 1.81  &  &  & &&\\
			\textbf{Ge$_{(B_{1})}$}  & 0.53   & \textbf{CsGeCl$_3$}& 1.11&1.11\cite{Radha2018}&2.01$^a$, 2.16$^b$& 2.10\cite{Bouhmaidi2022}, 3.67*\cite{Schwarz1996}\\
			\textbf{Cl$_{(X_{1})}$}  & 1.67  &&  & &&\\
			\hline
			\textbf{Rb$_{(A_{2})}$} & 1.52  & & & &&\\
			\textbf{Sn$_{(B_{2})}$} & 0.69  & \textbf{RbSnBr$_3$} & 0.89 &0.84\cite{Li2021}&2.39$^a$, 2.68$^b$&1.10\cite{Sabine2016}, 2.93\cite{Saikia2022}\\
			\textbf{Br$_{(X_{2})}$} & 1.82  & & &&&\\
			\hline
		\end{tabular*}
		Note: a = GGA, b = mGGA, * = experiment
	\end{table*}
	
	\par
\begin{figure}[hbt!]
		\small
		\includegraphics[height=10cm]{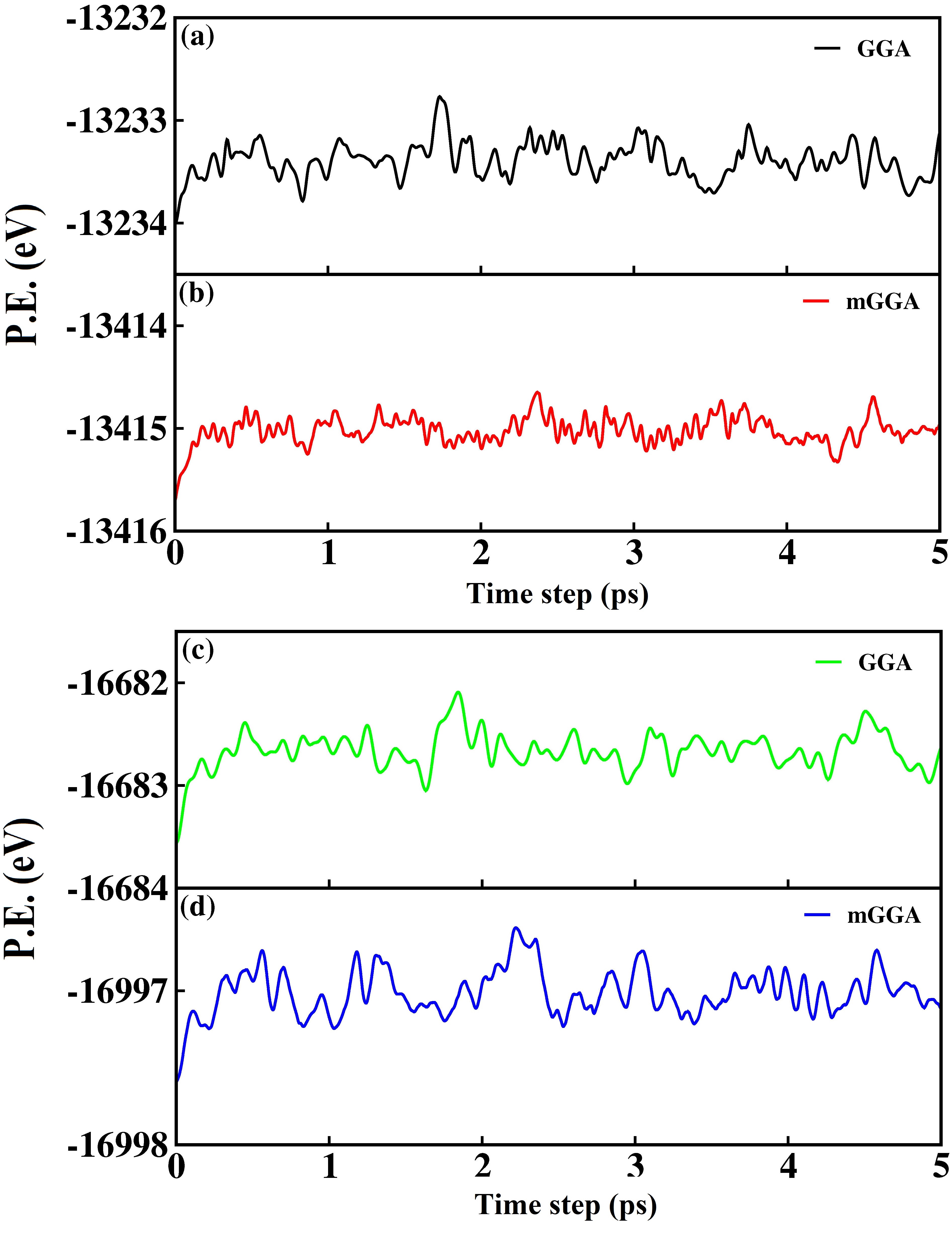}
		\caption{Molecular dynamic simulations of bulk CsGeCl$_3$- Fig.(a\&b) and RbSnBr$_3$- Fig.(c\&d) using GGA and mGGA}
		\label{Fig.2BulkMD}
	\end{figure}

 To validate and analyse atoms and molecular mobility, we conducted Molecular Dynamics (MD) simulations within a canonical ensemble employing Nose Hoover thermostat as implemented in QuantumATK's code\cite{Lalengmawia2025}. This technique has been utilized since the 1960s and is still a reliable tool to study the results of other classical methods, offering several uses in the fields of biology, biophysics, and nanotechnology\cite{Bonadeo1992}. Fig.\ref{Fig.2BulkMD} shows the development of potential energies (PE) for the examined eco-friendly perovskites up to 5ps time increments. Herein, we used the canonical ensemble based on nVT. To get understandable findings in the evolution of the energies, the number of particles, volumes, and temperature were kept constant during these simulations run at 300K. Since the simulated profiles of the PE indicate almost linear variations, this indicates that the systems are thermally stable. The calculated bulk's elastic properties to check the mechanical criterion and nature of the materials are kept in S2. Along with these, the study of charge transfer and the nature of bonding using EDD and ELF are kept in S3 and S4.
	
	\par
\begin{figure}[hbt!]
		\small
		\includegraphics[height=5.8cm]{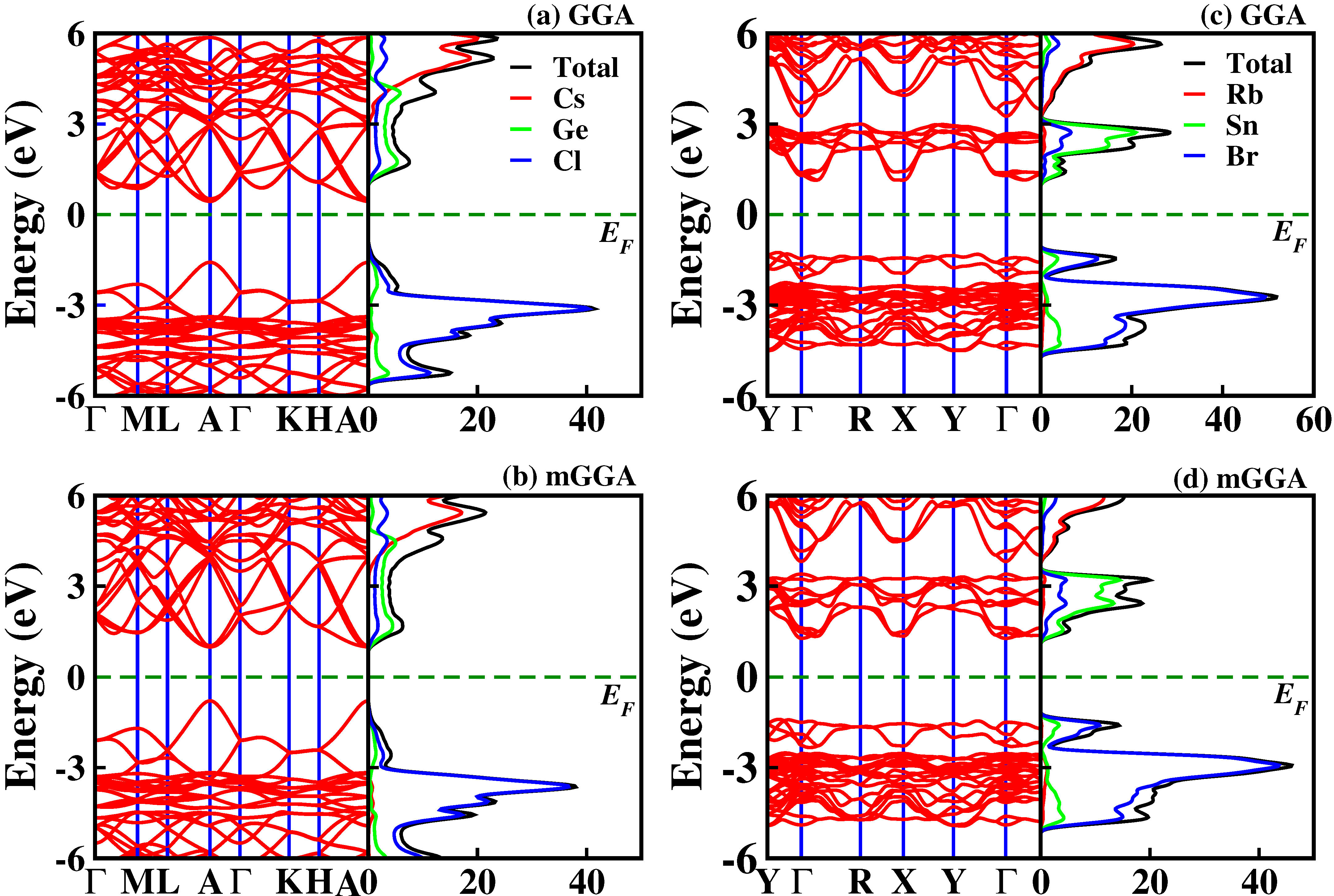}
		\caption{Band structures and DOS of CsGeCl$_3$ Fig.(a \& b) and RbSnBr$_3$ Fig.(c \& d).}
		\label{Fig.3BANDDOS}
	\end{figure}

 Examining a material's electronic properties remains a crucial parameter while studying the atomic-level interactions within a compound\cite{Celestine2024}. For the investigated systems, we employed two different exchange functionals, namely PBE-GGA as well as mGGA, to provide better accuracy. Utilizing these, we obtained a semiconducting direct band gap for CsGeCl$_3$ with $\Gamma$-M-L-A-$\Gamma$-K-H-A as the high symmetry points where the conduction band minimum (CBM) and valence band maximum (VBM) lie on the same high symmetry A-point. While an indirect band gap for RbSnBr$_3$ has been observed, which comprises of Y-$\Gamma$-R-X-Y-$\Gamma$ but the CBM and VBM do not lie on the same energy levels. These findings are reported in Table \ref{Table 1} and Fig.\ref{Fig.3BANDDOS}, taking the energy band range from -6 to +6 eV with zero-point energy as the Fermi energy (E$_F$). From the acquired band gaps, we observed that for CsGeCl$_3$, mGGA gave wider band gaps than the GGA outputs\cite{Wickramaratne2024}. Likewise, for RbSnBr$_3$, the GGA band gap is relatively lower than the mGGA result. This results in the addition of kinetic energy density, which helps in collecting more physical and accurate information for mGGA as compared to GGA. Fig.\ref{Fig.3BANDDOS} also depicts the density of states (DOS) for the studied compounds under GGA and mGGA. This study demonstrates how much each atomic orbital contributes to the exchange of photons between the valence and conduction bands. The valence orbital of each element has been considered. Herein, we have taken the energy range from -6 to +6 eV and amplitude up to 50 states/eV for both CsGeCl$_3$ (GGA and mGGA) and RbSnBr$_3$ (mGGA), but for RbSnBr$_3$'s GGA we have taken up to 60 states/eV due to higher atomic orbital's contribution. From the figures, we observed that CsGeCl$_3$ under both functionals, Cs-atom contributed at the high-energy levels while Ge- and Cl- atoms dominate near the Fermi-line, where the most dominant Ge-atom amplitudes reach up to 7.56 states/eV at 1.73 eV for GGA and 6.85 states/eV at 1.61 eV for mGGA in the conduction band. While in the valence band, the Cl-atom is the highest contributor whose amplitude reaches up to 41.26 states/eV at -3.17 eV for GGA and reaches up to 36.78 states/eV at -3.59 eV for mGGA. Likewise, for RbSnBr$_3$ under both functionals, we have observed that the contribution of the Sn-atom dominates in the conduction band, with its amplitude reaching up to 28.23 states/eV at 2.67 eV for GGA, while reaching around 19.52 states/eV for mGGA. Furthermore, in the valence band, we also observed that the contribution of Br-atom is the most with 48.88 states/eV at -2.74 eV for GGA and 46.15 states/eV at -3.04 eV for mGGA. Our findings reveal that CsGeCl$_3$ has a better transport range of photon energy from valence to conduction bands, while RbSnBr$_3$ gives higher atomic contributions.
	
	\par Determining a material's optical properties is crucial to understand its electronic configurations and how well it responds to electromagnetic radiation (light waves) for them to be used in various optoelectronic and photovoltaic applications\cite{Renthlei2023}. To comprehend the systems' responses to solar energy, we have calculated various optical properties, such as dielectric constants $\epsilon(\omega)$, absorption amount $\alpha(\omega)$, and refractive indices $\eta(\omega)$ at photon energies up to +10 eV. Due to the asymmetry of the understudied compounds, the optical responses for CsGeCl$_3$ are obtained along xx- and zz- directions, whereas those for RbSnBr$_3$ are obtained along xx-, yy-, and zz- directions. The expressions to obtain the complex dielectric functions, absorption coefficients are kept in the supplementary file (ESI). The real $\epsilon_1(\omega)$ and imaginary $\epsilon_2(\omega)$ profiles of the bulk systems' dielectric constants are depicted and discussed in S5. The absorption coefficients of the bulk systems are illustrated in Fig.\ref{Fig.9All-absorp} where the regular lines represent GGA and mGGA with the dotted lines. Herein, for CsGeCl$_3$, a sharp onset of 2.5 eV for both functionals has been observed, where the peaks fall within the region of vis-UV region, having the highest peak reaching up to $\alpha \sim$5.87$\times$10$^5$ cm$^{-1}$ (mGGA), making it relevant for optoelectronic and various photovoltaic applications. Additionally, for RbSnBr$_3$, the absorption onsets are different for each axis. As compared to CsGeCl$_3$, it has almost the same onset at around 2.5 eV along the xx-axis. While the peaks of all the axes fall within the vis-UV region, they are observed to have lower absorption coefficients, with the highest peak reaching up to $\alpha \sim$3.79$\times$10$^5$ cm$^{-1}$ (mGGA), which is lower than CsGeCl$_3$. In conclusion, we can say that CsGeCl$_3$ absorbs strongly in the higher vis-UV region while RbSnBr$_3$ absorbs more weakly in the lower visible energy region.

	\subsection{Surface Properties}
	
	\subsubsection{Structural \& stability, electronic and optical properties}
	\par To understand the surface properties of the materials, we have cleaved the surface of the under-investigated compounds from their respective bulk systems under the functionals employed. The Miller indices of \textit{h}=\textit{k}=0, \textit{l}=1 have been configured to define the surfaces. The in-plane surface lattices were fixed at
	\begin{equation}
		\begin{split}
			\nu_1=1 \ \mu_1 + 0 \ \mu_2 \\
			\nu_2=0 \ \mu_1 + 1 \ \mu_2 \\
			\label{Eq2}
		\end{split}
	\end{equation} 
	
	\par where $\nu_1$ and $\nu_2$ are the lattice vectors along \textit{a}- and \textit{b}-directions with $\mu_1$ and $\mu_2$ being the fundamental basis vectors in the surface cut plane. Perpendicular to these vectors, we have an out-of-plane cell vector, $\nu_3$, which often refers to the \textit{c}-direction, which defines the periodicity and vacuum of the surface. The parameters configured for CsGeCl$_3$ and RbSnBr$_3$ along c-direction are given in Table \ref{Table 2}
	
	\begin{table}[hbt!]
		\small
		\caption{\ Parameters given for surface slab cut (in \AA)} 
		\label{Table 2}
		\begin{tabular*}{0.48\textwidth}{@{\extracolsep{\fill}}l|l|l}
			\hline
			& CsGeCl$_3$ &RbSnBr$_3$\\
			\hline
			Top Vacuum& 0 & 16.46\\
			Thickness& 28.61 & 36.38\\
			Bottom Vacuum & 10.22  & 0\\
			Selected atom placement, c& 0.74 & 0.69 \\
			\hline
		\end{tabular*}
	\end{table} 
	
	\par
\begin{figure}
		\small
		\centering
		\includegraphics[height=3.7cm]{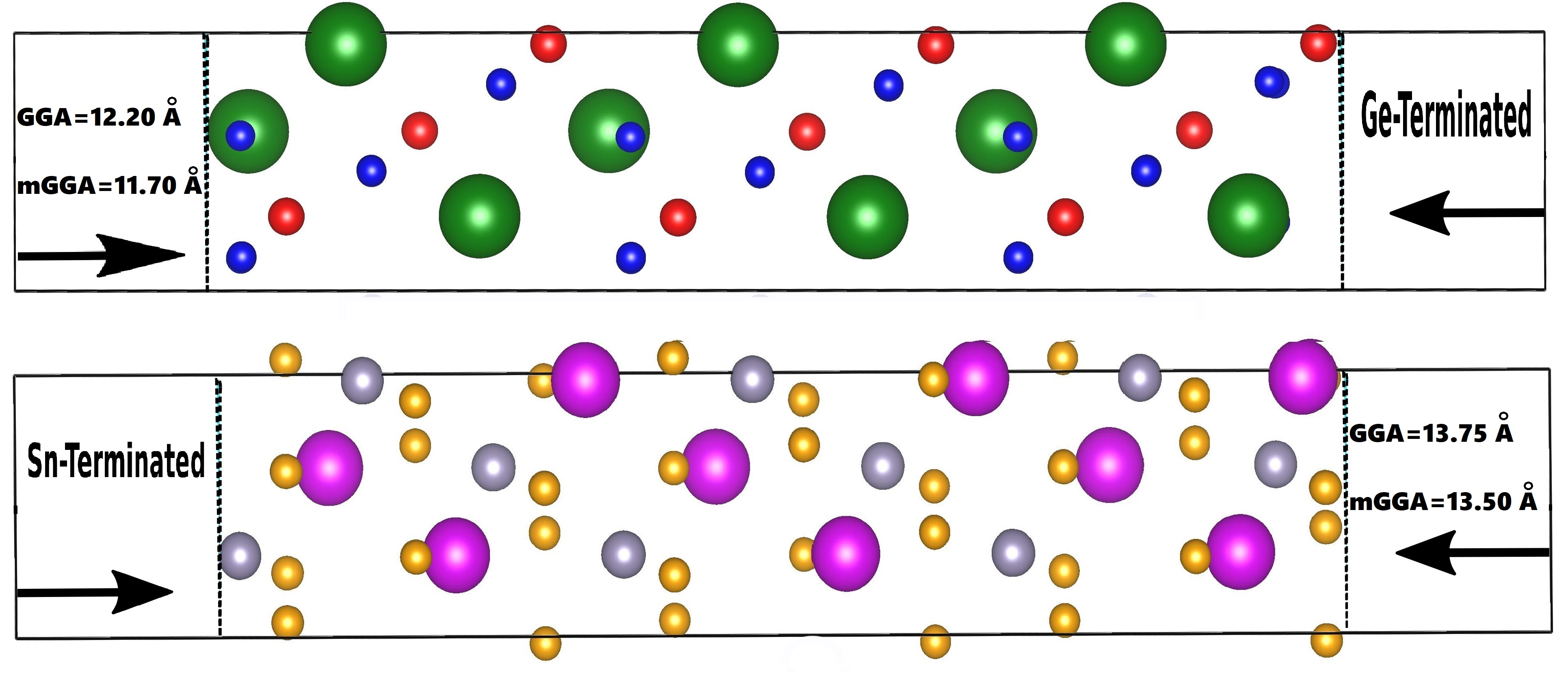}
		\caption{Schematic surface structures of CsGeCl$_3$ (001) and RbSnBr$_3$ (001)}
		\label{Fig.4surface-structures}
	\end{figure}

 The extended direction of the slab cut from the bulk systems is tabulated in Table \ref{Table 2}. We have given 10.22\AA \ and 16.46\AA \ as the top and bottom vacuum for the investigated surface slabs to prevent interaction between the periodic images within the non-periodic \textit{c}-direction. Implying this range of slabs vacuum spacing is acceptable by the existing literature reported by Grancini et al.\cite{Grancini2017}, Sun et al.\cite{Sun2013}, and Khomyakov et al\cite{Khomyakov2009}. Consequently, to avoid spurious interactions, the thickness specified for both slab systems is found to be adequate for surface analysis. Furthermore, the \textit{c}-value indicates the relative vertical position, whose value ranges between 0$\rightarrow$1, is the key atom in the simulation cell. Our values obtained given in the table indicate the atoms are located towards the top portion of the slabs. Fig.\ref{Fig.4surface-structures} shows the obtained surface models' Top(T)- and Bottom(B)- terminations for both the studied slabs. For CsGeCl$_3$, we obtained the Ge- and CsCl- terminations. While RbBr- and Sn- terminations for RbSnbr$_3$. Surface slabs optimized for mGGA have a significantly lower thickness than their GGA counterparts. This reduction is due to the enhanced electron localization in mGGA, which improves surface bonding and inhibits the outward relaxation that GGA often overestimates.
	
	\par Fig.\ref{Fig.4surface-structures}, provides the atomic arrangements and repetitions of the atoms in each surface slab. Fig.S7 provides a systematic presentation of all the structural parameters for each completely relaxed surface slab model, including the in-plane lattice constants, band gaps, and volumes. The relaxed atomic displacements (d$_f-d_i$ = d$_{relaxed}$ - d$_{unrelaxed}$) for each atom have been observed, and we report that for the compound CsGeCl$_3$, all atoms are in an inward relaxation under GGA, while parts of the Cs and Ge atoms are in outward relaxations for mGGA. But for RbSnBr$_3$, except for the first 4 atoms (Sn, Br, Br, and Br), all are directed to inwards relaxation, which is also observed under mGGA except for the first Br-atom. The values of these are shown in AD1-AD4. 
	
	\par
\begin{figure}[hbt!]
		\centering
		\includegraphics[height=5cm]{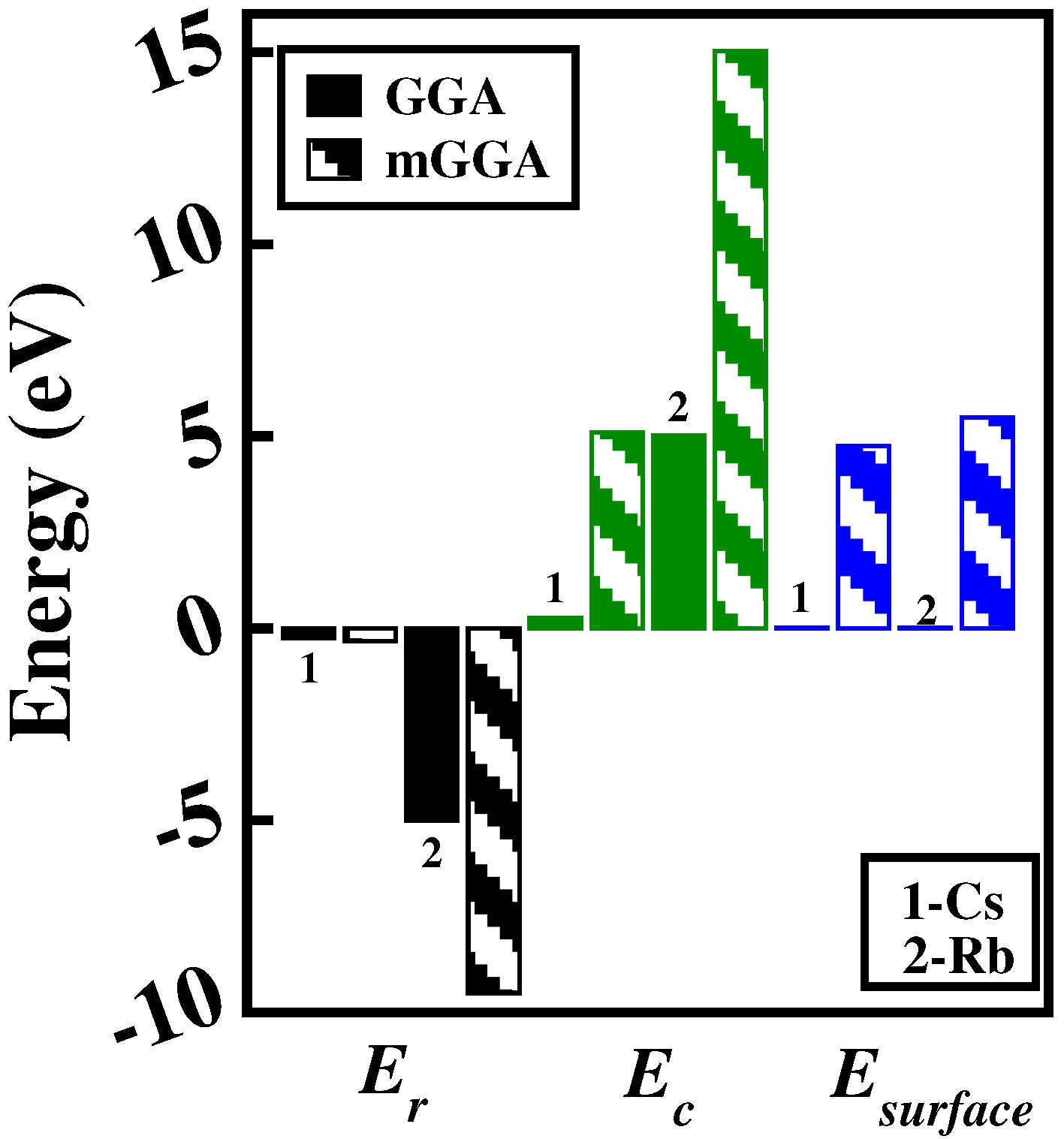}
		\caption{Relaxation energies (E$_r$), Cleavage energies (E$_c$), and Surface energies (E$_{surface}$) of CsGeCl$_3$ (001) and RbSnBr$_3$ (001)}
		\label{Fig.5Energies}
	\end{figure}

 For a surface slab model to be energetically stable, we have performed the surface energy calculations  (E$_{surface}$), using the two key components:\cite{Zosiamliana2025} (a) the relaxation per unit surface area (E$_r$), which measures the energy transition that takes place when a surface slab goes from unrelaxed to a relaxed state, and (b) the cleavage energy per unit surface energy (E$_c$), which signifies the required energy to cleave surfaces from the bulk material. The obtained values of E$_r$ and E$_c$ are illustrated in Fig.\ref{Fig.5Energies}. Whilst the mathematical expressions for calculating these values are given as follows:\cite{Eglitis2022, Lalroliana2023}
	
	\begin{equation}
		\begin{split}
			E_r(X) = \dfrac{1}{2A} \times \Bigg( E_{slab}^{relaxed} (X) - E_{slab}^{unrelaxed} (X)\Bigg)\\
			E_c(X) = \dfrac{1}{2A} \times \Bigg( E_{slab}^{relaxed} (X) - n.E_{bulk}\Bigg)\\
			E_{surface}(X) = E_r(X) + E_c(X)
		\end{split}
		\label{Eq.3}
	\end{equation}
	
	\par Apparently, the obtained negative values of E$_r$ for all the investigated surface slab models assuredly mark the relaxation course to be energetically favourable, promoting a net energy release and moving the system in a manner of a more stable configuration. This means that atoms at the surface normally adjust their positions to lessen and minimize the surface stress and total energy. Consequently, the relatively small values of E$_r$ indicate that the arrangements are refined and primarily involve minor local displacements instead of extensive structural rearrangements. This type of behaviour indicates the lack of noticeable surface reconstruction, which usually shows up as atomic rearrangement and substantially greater relaxation energies. Therefore, with modest surface relaxations, energy stabilisation is achieved while the surface slabs maintain their overall structural integrity. As shown in Fig.\ref{Fig.5Energies}, relaxation energies (E$_r$) obtained by CsGeCl$_3$ are more stable, with less decay time and can operate better at high temperatures than RbSnBr$_3$. On the other hand, RbSnBr$_3$ are superior towards energy dissipation and adaptability, in storing energy and absorption of shock and vibrations. From the principles of fundamental surface physics\cite{Bechstedt2003}, E$_c$ indicates the energy required to cleave two new surfaces from the bulk system. Since the process of breaking the chemical bond involves endothermic action, which implies a positive E$_c$ value and absorption of energy by the material is obligatory to counteract the cohesive force within the system. Therefore, our results show that the compounds under GGA tend to have weaker chemical bondings, which makes them easier to separate into layers and helps manufacturing easily. While compounds under mGGA have stronger bonds, resulting in more resistance to breaking. Utilizing the first two expressions from Eq.\ref{Eq.3}, we acquired the values of E$_r$ and E$_c$. With these, we finally calculate and analyse the E$_{surface}$ of the surface slab models. From Fig.\ref{Fig.5Energies} it is determined that the compounds under the GGA functional employed are more energetically stable due to their lower reactivity with atoms at the surface are tightly bonded. The values of the calculated relaxed and unrelaxed energies of the surface slabs are kept in S8. Along with these, the calculated MD simulations, EDD, and ELF for these surfaces under GGA and mGGA are kept and discussed in S9, S10, and S11.
	
	\clearpage
	\par
\begin{figure}[htbp]
		\centering
		\includegraphics[height=8cm]{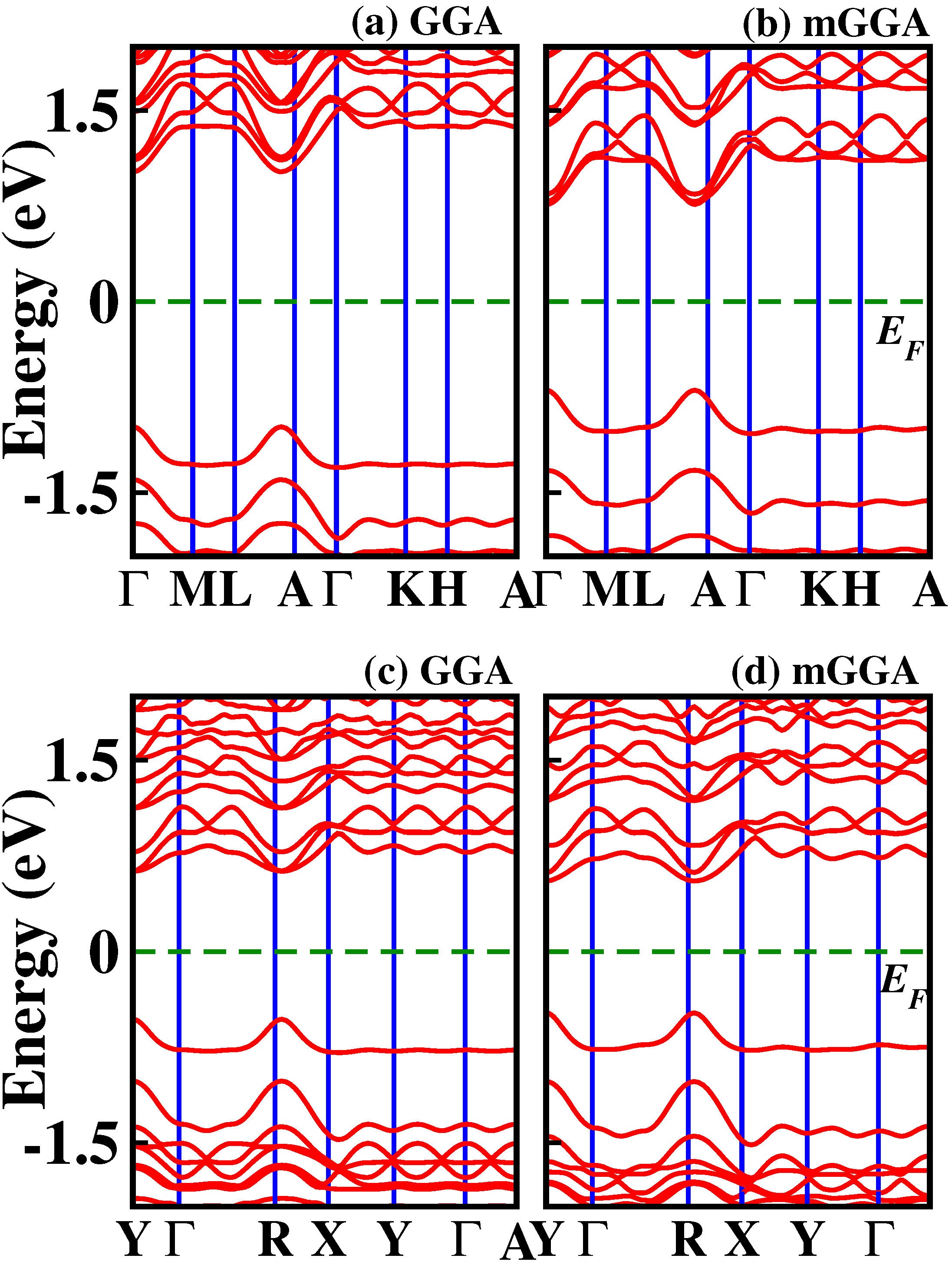}
		\caption{Electronic band structures of CsGeCl$_3$ (001) Fig.(a \& b) and RbSnBr$_3$ (001) Fig.(c \& d)}
		\label{Fig.6slabelectronics}
	\end{figure}

 Analysing the electronic structure of any surface model is a vital way to comprehend a material's interaction with its surroundings and how it influences the key properties, including optical attributes and much more. Fig.\ref{Fig.6slabelectronics} shows the band dispersion of the under-investigated surface slabs CsGeCl$_3$ (001) and RbSnBr$_3$ (001), respectively. In this instance, we can clearly see that the employed functionals: GGA and mGGA play vital roles in our obtained band structures. The GGA approach is known to underestimate the band gap as it lacks the derivative discontinuity because it relies on local electron density and its gradient, which proves inadequate for accurately capturing essential electron properties, including the fundamental band gap. The mGGA approach is influenced not just by electron density and its gradient, but also by the kinetic energy density, which is essential for effectively capturing non-local effects that are vital for precise band gap calculations. As depicted in Fig.\ref{Fig.6slabelectronics}, we observed that despite the predicted underestimation of band gaps by GGA, the values obtained are wider than the mGGA results, which is unusual. In most cases, mGGA with its kinetic energy density and functional's curvature on the conduction and valence bands increases the band gaps. In particular, our calculated mGGA band gaps, which lack derivative discontinuity, produce smaller band gaps than the GGA's. The density of states (DOS), which discusses about the contribution of valence orbital for each atom are reported and kept in S12 The complex dielectric functions ($\epsilon_1(\omega)$ and $\epsilon_2(\omega)$) which determines the optical properties of the surface slabs are given in S13. From the imaginary part discussed in S13(b \& d), we obtained the absorption coefficients of the studied slab models depicted in Fig.\ref{Fig.9All-absorp}. Herein, under both the functionals, a sharp onset of 2.12 eV for CsGeCl$_3$ has been observed with the highest peak along the zz-axis reaching up to $\alpha$ $\sim$ 1.65$\times$10$^5$ cm$^{-1}$ whose value falls within the vis-UV region. Furthermore, for RbSnBr$_3$, a sharp onset of around 1.23 eV with the highest coefficient reaches up to $\alpha$ $\sim$0.38$\times$10$^5$ cm$^{-1}$ with the region of vis-UV has been observed. Therefore, following the trend of the bulk systems, we can conclude that CsGeCl$_3$ has a higher $\alpha$-coefficient in the higher vis-UV part, while RbSnBr$_3$ has lower absorption in the lower visible energy section. Detailed results and discussion of elastic and mechanical properties are kept in S14.

	\subsection{Heterostructure Properties}
	\subsubsection{Structural \& stability, electronic and optical properties}
	
	\par This section highlights the construction of a heterostructure employing two surface slab model semiconductors based on optimal energy band alignment to produce an energy band offset between them. It facilitates the effective transport of electrons and holes between the valence and conduction bands, thus enhancing the charge carrier separation and migration efficiency. Using the coincidence site lattice approach\cite{Hinz1995} and an nV$_1$ + mV$_2$ grid size, the two in-plane lattices are matched to create an interface model. A total number of 90 atoms, 45 atoms from both surface slabs, are present in our constructed heterojunction. For which we combined the T-terminated Ge-atoms of CsGeCl$_3$ (001) and B-terminated Sn-atoms of RbSnBr$_3$ (001), avoiding possible and high percentage mismatch with other terminations. Due to the mismatch percentage of 5\%, which is regarded as excessive\cite{Zheng2015, Attia2019}, our lattice mismatch percentages of 4.44\% and 2.70\%, obtained from Table \ref{Table 3}, for \textit{a} and \textit{b} are considered to be acceptable. Likewise, for the mean absolute strains- the first, second, and mean surface strain yields highly favourable, low percentages of 2.21\%, 2.13\%, and 1.09\%, which is excellent for an ideal heterostructure. Consequently, the obtained heterostructure can be considered to be periodic along the xy-plane (lateral) with z-direction as the out-of-plane or polar direction (non-lateral). The approximate lattice constant ratio was taken in order to determine the anisotropic-periodic lengths of the crystal structures. Our \textit{a/b}-values indicate that the crystals formed were deviated and slightly distorted, indicating an asymmetrically arranged structure.
	
	\par
\begin{figure*}
		\centering
		\includegraphics[height=2.5cm]{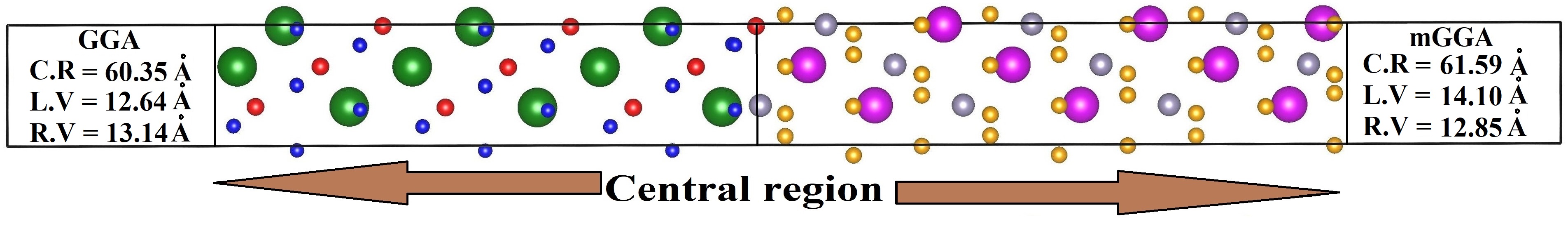}
		\caption{Schematic figure of heterostructure- CsGeCl$_3$/RbSnBr$_3$ using GGA and mGGA}
		\label{Fig.7Interfacestructure}
	\end{figure*}

 From Fig.\ref{Fig.7Interfacestructure}, we can see the thickness difference of the constructed heterostuctures for GGA and mGGA. The central region (C.R.) under mGGA gives slightly higher thickness than the GGA's ($\sim$1\AA). As GGA compressed and relaxed the atoms inward, mGGA corrects the relaxation with better electron localization, which increases the effective repulsiveness. Thus, it results in relaxing the atoms outward. Following this trend, with two non-identical and asymmetric materials, the left vacuums (L.V) and right vacuums (R.V) also differ in the functionals employed. These outcomes present that our constructed heterostructures are in asymmetric arrangements exhibiting a built-in dipole at the interface, indicating a highly polarizable nature, which gives a green flag to proceed with our work. Detailed heterostructural parameters and MD simulations for the newly constructed heterostructures to check the energy fluctuations for thermal stability are kept in S15 and S16. 
	
	\par The energy required to create a new interface region by increasing the amount of atoms at the interface is known as the interface energy ($\gamma$) of the heterophase junction. It can also be deciphered as the sum of the surplus energies of all atoms around that interface. These extra energies define the difference between an atom's real energy and its energy in the same deformative state, which is in an ideal homophase crystal. The $\gamma$-values for the interface models under GGA and mGGA are tabulated in Table \ref{Table 3}. The positive $\gamma$-value under GGA indicates an energy cost, meaning work must be done to create the interface. However, a negative $\gamma$-value has been observed for the interface under mGGA, indicating a release of energy while forming the interface. This negative value may arise due to the presence of chemical inhomogeneity and structural distortions in the interface. Interfacial energies with positive and negative $\gamma$-values have also been reported by P.Gumbsch and M. S. Daw \cite{Gumbsch1991}. Additionally, there have been findings where interfacial energies are observed to be negative. Recently, Saito et al.\cite{Saito2023} have determined negative $\gamma$-values in their various polymer interface studies. These signify that work outputs are in agreement with the existing data or literature. The charge distribution and localizations of the heterostructures under GGA and mGGA are discussed in S17 and S18.
	
	\begin{figure}[hbt!]
		\centering
		\includegraphics[height=4.8cm]{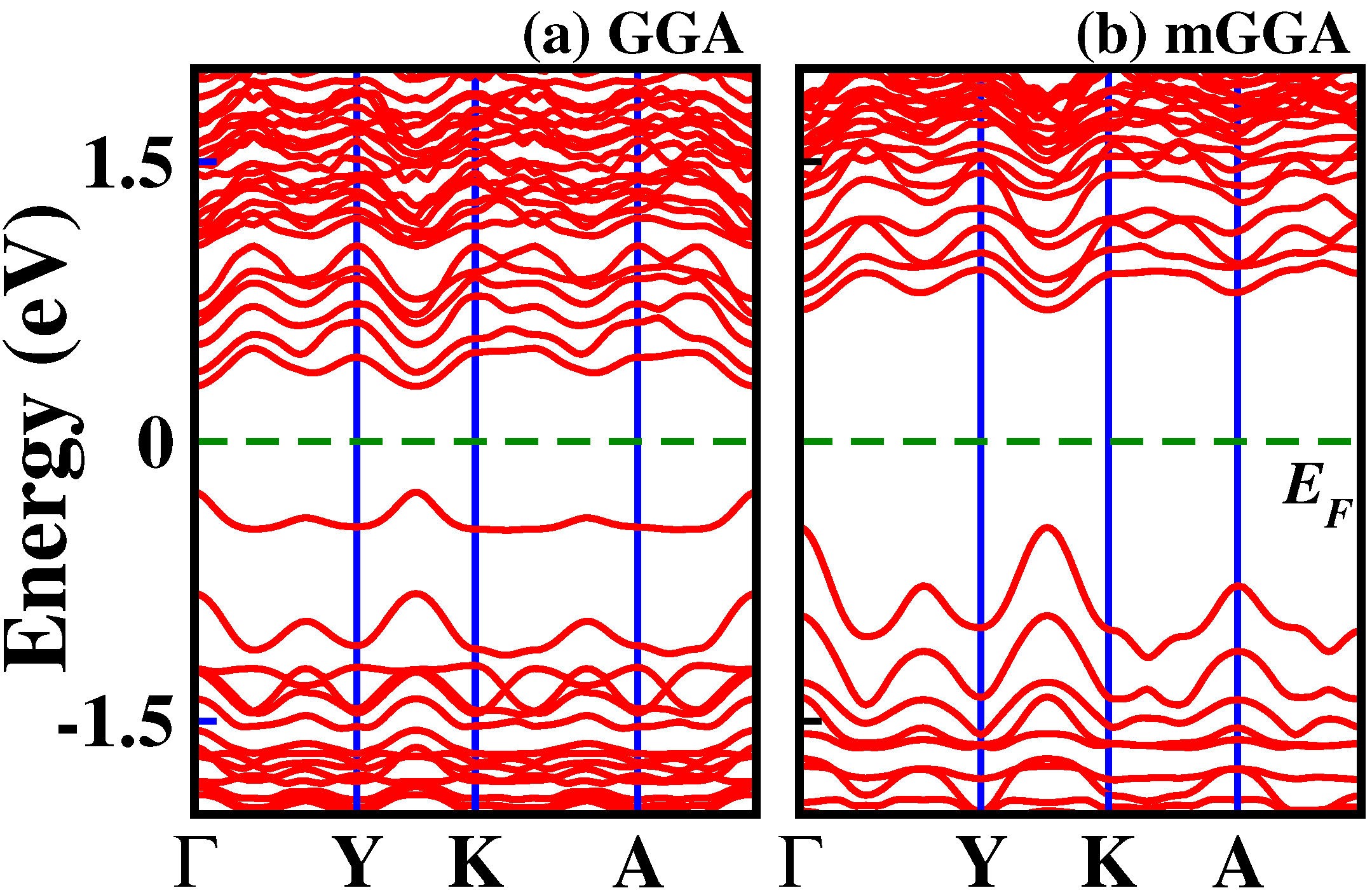}
		\caption{Electronic band structures of heterostructure- CsGeCl$_3$/RbSnBr$_3$. Fig.(a)GGA and Fig.(b)mGGA}
		\label{Fig.8interelectronic}
	\end{figure}

	\begin{figure*}
		\centering
		\includegraphics[height=4cm]{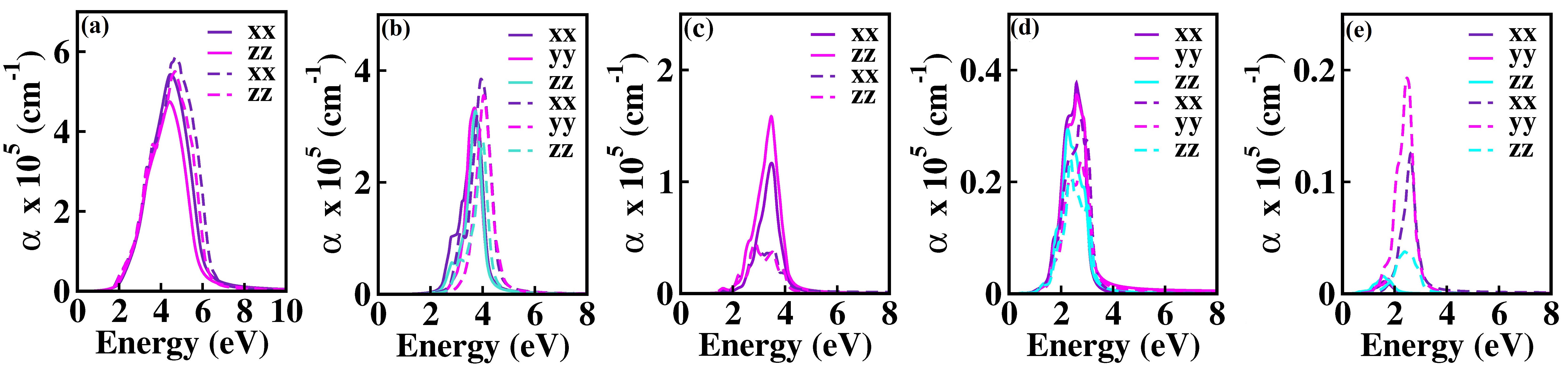}
		\caption{  Absorption coefficients of: bulk- (a)CsGeCl$_3$, (b)RbSnBr$_3$: slabs- (c)CsGeCl$_3$, (d)RbSnBr$_3$ and (e)heterophase}
		\label{Fig.9All-absorp}
	\end{figure*}
	\par The electronic properties of the CsGeCl$_3$/RbSnBr$_3$ heterostructure have been examined through the incorporation of GGA and mGGA functionals. As shown in Fig.\ref{Fig.8interelectronic}, we have obtained semiconducting heterostructures under the influence of the said functionals employed. Widening of band gaps from GGA$\rightarrow$mGGA has been observed in our constructed heterophase, indicating possible transfer of charge between the valence and conduction bands. In addition to that, the direct band gaps observed in the band structures due to the direct alignment of the VBM and CBM allow better transition of electrons, making them ideal towards optoelectronic applications. The VB is dominated by Cl-p and Ge-p states from the CsGeCl$_3$ surface layer, and the CB is predominated by the Sn-p and Br-p states of the RbSnBr$_3$ layer, indicating localization of holes and electrons are on different materials, confirming a Type II (staggered) band alignment, promoting efficient charge separation, which is better in the field of photovoltaic and optoelectronics applications. The atomic orbital contributions, density of states, are kept in Fig.S19. The optical characteristics of the constructed heterostructures have been examined and discussed. The obtained real ($\epsilon_1(\omega)$ and imaginary $\epsilon_2(\omega)$) parts of the dielectric constants are discussed and kept in Fig.S20. Based on the imaginary parts of the dielectric constants, we have the absorption coefficients ($\alpha$) for the interface system under investigation. As seen in Fig.\ref{Fig.9All-absorp}, we report that all the $\alpha$-peaks along x-, y-, and z-directions reach up to 0.02$\times$10$^5$cm$^{-1}$ under GGA, while higher peaks are observed for mGGA, which are in the range of 0.03-0.19$\times$10$^5$cm$^{-1}$. Our obtained $\alpha$-values are suitable for an efficient interface absorber, showing promising results to be used in PV or optoelectronic applications. Furthermore, the presence of absorption peaks in the low visible range supports the Type II band offset that permits interfacial charge-transfer transitions by confirming that the CsGeCl$_3$/RbSnBr$_3$ interface maintains its semiconducting nature with an effective visible-light response.

	\begin{table*}[hbt!]
		\centering
		\caption{\ Calculated lattice ratio(in \AA), mismatch (in \%), and interfacial energies ($\gamma$)} 
		\label{Table 3}
		\begin{tabular*}{\textwidth}{@{\extracolsep{\fill}}l|l|l|l|l|l|l|l}
			\hline
			\textbf{Slabs} & Orientation & Xc & a/b & mismatch (x100)& Interface&Xc&$\gamma$ \\
			\hline
			\textbf{CsGeCl$_3$} & (001)  & GGA & 0.96 & 0.34/7.65&\textbf{CsGeCl$_3$/RbSnBr$_3$}&GGA&0.11\\
			& & mGGA & 0.96 & &&&\\
			\textbf{RbSnBr$_3$ } & (001)  & GGA & 0.97 & 0.21/7.78&\textbf{CsGeCl$_3$/RbSnBr$_3$}&mGGA&-0.001\\
			& & mGGA & 0.94 & &&&\\
			\hline
		\end{tabular*}
	\end{table*}

	\subsection{Piezoelectric Coefficients and Electromechanical Coupling Constants}
	\par Due to their potential for harnessing renewable energy, piezoelectric features have attracted a lot of attention. A piezoelectric material generates electrical energy by polarizing positive and negative charges in response to external forces. Non-centrosymmetric materials respond piezoelectrically \cite{Celestine2025}. Thus, higher polarization results in a higher piezo-response. To meet the demand for energy storage and generation, both natural and man-made piezoelectric materials are being explored and developed. For this, less toxic and environmentally-friendly compounds are in high demand\cite{Celestine2024}. Our proposed compounds, which are Pb-free systems, can be promising candidates in this field. 
	
	\par The primary goal of analyzing the piezoelectric characteristics of CsGeCl$_3$ and RbSnBr$_3$' is to determine their tensors in their bulk, surface slab and heterostructure forms. Upon applying external pressure, we compute the piezoelectric tensors. Just a handful of data for bulk' piezo-response can be found \cite{Celestine2024}. However, no other piezo-data has been reported for the slabs and heterophases. Since the Sn- atom is larger than the Ge- atoms, these reflect a polar mismatch local strain gradient generation, which may enhance the piezoelectric coefficients in the constructed heterostructure. In addition to piezoelectric tensors, we computed the electromechanical coupling constants to validate the efficacy and electric rate of conversion for the lead-free perovskites, utilizing elastic constants, dielectric constants, and free space permittivity. Without the external fields, the total macroscopic polarization (P) is calculated by adding the spontaneous polarization (P$_{eq}$) of the equilibrium structure and the strain-dependent piezoelectric polarization (P$_p$), whose expression is given as
	
	\begin{equation}
		P = P_{eq} + P_p
		\label{Eq 4}
	\end{equation}
	
	Therefore, the piezoelectric tensors can be acquired as 
	
	\begin{equation}
		\gamma_{\delta\alpha} = \dfrac{\Delta P_\delta}{\Delta \epsilon_\alpha}
		\label{Eq 5}
	\end{equation} 
	
	\par Herein, with the employed Berry-phase approximation, we obtained the polarization (P). The piezoelectric coefficient ($\gamma_{\delta\alpha}$) can be obtained using the change in polarization by the applied strain, employing the finite difference technique.
	
	\par An alternate method to acquire the piezoelectric tensors can be done by setting up two terms:
	\par (1) The electronic response to strain (e$_{ij}$) also known as the clamped-ion;
	\par (2) The statement which describes the impact of pressure on polarization.
	\par Therefore, the expression for e$_{ij}$ is given as
	
	\begin{equation}
		e_{ij} = e_{ij}(0) + \dfrac{4eZ^*}{\sqrt{3a^2}} \dfrac{du}{d\epsilon_\alpha}
		\label{Eq 6}
	\end{equation}
	
	\par where \textit{i} and \textit{j} represent the applied current and strain directions, \textit{e} illustrates the electronic charge while Z $^*$ represents the Born effective charge. With \textit{a} as the lattice constants, \textit{u} being the interatomic distance, and $\epsilon_\alpha$ be the macroscopic applied strain.
	
	\begin{table*}[hbt!]
		\small
		\caption{\ Computed piezoelectric constants of the surface slabs CsGeCl$_3$ and RbSnBr$_3$ using GGA and mGGA (all in C/m$^{2}$)}
		\label{Table 4}
		\begin{tabular*}{\textwidth}{@{\extracolsep{\fill}}|l|l|l|l|l|l|l|l|l|l|l|l|}
			\hline
			&&&&&&\textbf{CsGeCl$_3$}&&&&&\\
			\hline
			\textbf{GGA}&  & &&&  &\textbf{mGGA}&&&&&\\
			\hline
			e$_{11}$ & \ 7.91e$^{-01}$ &e$_{21}$& \ 1.24e$^{+00}$ &e$_{31}$& 4.65e$^{-02}$  	&e$_{11}$ & \ 5.08e$^{-01}$ &e$_{21}$& -9.72e$^{-02}$ &e$_{31}$&\ 3.96e$^{-02}$\\
			e$_{12}$ &  \ 8.84e$^{-01}$ &e$_{22}$& -1.08e$^{+00}$&e$_{32}$& 4.97e$^{-02}$ &	e$_{12}$ & -1.09e$^{+00}$ &e$_{22}$& -1.06e$^{+00}$&e$_{32}$&  \ 5.45e$^{-02}$\\
			
			e$_{13}$&-1.36e$^{+00}$& e$_{23}$& \ 4.82e$^{-01}$ &e$_{33}$&1.68e$^{-02}$  &e$_{13}$&-2.24e$^{+00}$& e$_{23}$&\ 2.08e$^{+00}$ &e$_{33}$&\ 3.09e$^{-02}$\\		
			e$_{14}$ & \ 9.93e$^{-01}$ &e$_{24}$& -2.19e$^{+00}$&e$_{34}$& 1.77e$^{-02}$ &e$_{14}$ & \ 3.22e$^{-01}$ &e$_{24}$& -2.09e$^{+00}$&e$_{34}$& \ 1.30e$^{-02}$\\
			e$_{15}$ &  -4.25e$^{-02}$ &e$_{25}$& \ 1.21e$^{+00}$ &e$_{35}$&  5.02e$^{-02}$ &e$_{15}$ & -1.67e$^{+00}$ &e$_{25}$& \ 4.12e$^{-01}$ &e$_{35}$&-3.25e$^{-03}$\\
			e$_{16}$& \ 1.34e$^{+00}$ &e$_{26}$& -1.19e$^{+00}$ &e$_{36}$&6.09e$^{-03}$&e$_{16}$& -2.34e$^{+00}$ &e$_{26}$& \ 6.47e$^{-01}$  &e$_{36}$& \ 6.78e$^{-04}$\\
			\hline
			&&&&&&\textbf{RbSnBr$_3$}&&&&&\\
			\hline	
			
			\textbf{GGA}&  & &&&  &\textbf{mGGA}&&&&& \\
			\hline
			e$_{11}$ & \ 2.08e$^{-02}$ &e$_{21}$& \ 7.79e$^{-02}$ &e$_{31}$& -3.48e$^{-02}$&	e$_{11}$ & \ 2.55e$^{-02}$ &e$_{21}$&  \ 7.81e$^{-02}$ &e$_{31}$&-2.93e$^{-02}$\\
			e$_{12}$ &  -2.17e$^{-02}$ &e$_{22}$& -7.78e$^{-02}$&e$_{32}$& -3.49e$^{-02}$ &e$_{12}$ &  -1.96e$^{-02}$ &e$_{22}$& -7.69e$^{-02}$&e$_{32}$& -3.29e$^{-02}$\\
			
			e$_{13}$&-2.96e$^{-04}$& e$_{23}$&-3.06e$^{-05}$ &e$_{33}$&\ 3.69e$^{-03}$ &e$_{13}$&\ 2.21e$^{-02}$& e$_{23}$&-3.35e$^{-01}$ &e$_{33}$& \ 6.78e$^{-03}$\\		
			e$_{14}$ & -8.44e$^{-04}$ &e$_{24}$& -3.27e$^{-01}$&e$_{34}$& -3.43e$^{-04}$&e$_{14}$ & \ 2.21e$^{-02}$ &e$_{24}$& -3.35e$^{-01}$&e$_{34}$& \ 6.78e$^{-03}$\\
			e$_{15}$ & -2.89e$^{-01}$ &e$_{25}$& -1.53e$^{-01}$&e$_{35}$& -7.02e$^{-05}$ &e$_{15}$ & -2.92e$^{-01}$ &e$_{25}$& -1.39e$^{-01}$&e$_{35}$& -4.22e$^{-03}$\\
			e$_{16}$& \ 6.75e$^{-02}$ &e$_{26}$& -1.87e$^{-02}$&e$_{36}$& \ 2.16e$^{-05}$&e$_{16}$& \ 6.85e$^{-02}$ &e$_{26}$& -1.37e$^{-02}$&e$_{36}$& \ 3.49e$^{-03}$\\
			\hline
		\end{tabular*}
	\end{table*}

	\par The investigated Pb-free halide perovskites' piezo-tensors are generated along the directions: \textit{x, y,} and \textit{z} axes in six different strains: \textit{xx, yy, zz, yz, xz,} and \textit{xy}, respectively. The piezo-responses are therefore generated in Coloumb per square meter (C/m$^2$), taking the surface induced charge into account when an external force is applied to the material's surface. The piezo-tensors for the bulk systems under the employed functionals are tabulated and discussed in Table S6. For the surface slab-modeled systems, we have obtained 18 piezoelectric tensors for each under GGA and mGGA. In Table \ref{Table 4}, we observe that the responses generated by mGGA are slightly higher than the output of GGA, where \textit{e$_{16}$} = -2.34 C/m$^2$ (mGGA) and \textit{e$_{23}$} = -0.35 C/m$^2$ (mGGA) are the highest values obtained by CsGeCl$_3$ (001) and RbSnBr$_3$ (001), respectively. In general, our piezo-responses tabulated in Table \ref{Table 4} illustrate increasing values than their bulk systems. Moreover, these values are much higher than those of the response generated by Al-doped $\beta$-Ga$_2$O$_3$ (100) bilayer with -5.55e$^{-12}$ C/m$^2$ reported by Yu et.al\cite{Chen2025}. Furthermore, Alexandru and Sohrab\cite{Georgescu2019} reported a piezo-response of 0.1 C/m$^2$ on the surface of sapphire- Al$_2$0$_3$ (001), where the result is much lower than our obtained values. Additionally, Chen et.al \cite{Chen2024} studied vertically polarized heteroanionic 2D layered material- Sb$_2$TeSe$_2$ and reported piezoelectric coefficients of -0.79 C/m$^2$ from the first principles and around 4.54 pm/V ($\sim$ 4.54e$^{-12}$ C/m$^2$) using dual AC resonance tracking (DART) technique of PFM, which comes out to be less than our values. Implying our piezoelectric tensors are reasonable and these slabs are promising candidates for ferroelectric and piezoelectric applications. Finally, the heterostructure's piezoelectric coefficients acquired using GGA and mGGA are kept in Table \ref{Table 5}. Interestingly, the novel study of Pb-free HPs heterophase exhibits fair and excellent piezo-tensors. As expected, mGGA responses are higher than the GGA results due to the wider band gap and higher polarization. With \textit{e$_{35}$} = 2.32 C/m$^2$ (mGGA) and \textit{e$_{11}$} = -1.35 C/m$^2$ (GGA) are the highest piezo-coefficients from the functionals employed. Some of the obtained results are comparable to our surface slab piezo-tensors, indicating the heterostructure under study is capable of being utilize in the field of ferroelectric or piezoelectric systems. Consequently, our results are much higher than the response of the experimentally studied layered In$_2$Se$_3$/MoS$_2$ van der waals heterostructure with 17.5 pm/V ($\sim$ 17.5e$^{-12}$ C/m$^2$) by Yuan et.al\cite{Yuan2020}. Additionally, Rana et.al\cite{Rana2022} have unveiled an excellent piezoelectric response for FASnBr$_3$/PDMS polymer of around 50pm/V ($\sim$ 50e$^{-12}$ C/m$^2$), which is also lesser than our obtained tensor value. These results imply that our heterostructures are promising candidates towards storing and generating energy through piezoelectric methods.
	
	\begin{table}[hbt!]
		\small
		\caption{\ Computed piezoelectric constants of the heterostructure CsGeCl$_3$/RbSnBr$_3$ using GGA and mGGA (all in C/m$^{2}$)}
		\label{Table 5}
		\begin{tabular*}{0.48\textwidth}{@{\extracolsep{\fill}}|l|l|l|l|l|l|}
			\hline
			\textbf{GGA}&  & &&& \\
			\hline
			e$_{11}$ & -1.35e$^{+00}$ &e$_{21}$& \ 7.55e$^{-01}$ &e$_{31}$& \ 5.24e$^{-03}$  \\
			e$_{12}$ &  \ 7.19e$^{-01}$ &e$_{22}$& -8.36e$^{-01}$&e$_{32}$& \ 1.99e$^{-03}$\\
			
			e$_{13}$&-6.68e$^{-01}$& e$_{23}$& -5.72e$^{-01}$ &e$_{33}$&\ 1.62e$^{-02}$ \\		
			e$_{14}$ & -1.37e$^{-01}$ &e$_{24}$& -4.67e$^{-01}$&e$_{34}$& -6.29e$^{-03}$ \\
			e$_{15}$ & -3.47e$^{-01}$ &e$_{25}$& \ 8.32e$^{-01}$ &e$_{35}$& \ 2.53e$^{-02}$ \\
			e$_{16}$& -7.37e$^{-01}$ &e$_{26}$& \ 9.29e$^{-02}$ &e$_{36}$&\ 4.25e$^{-03}$\\
			\hline
			
			\textbf{mGGA}&  & &&&  \\
			\hline
			e$_{11}$ & -1.47e$^{+00}$ &e$_{21}$& \ 6.94e$^{-01}$ &e$_{31}$& \ 5.57e$^{-01}$\\
			e$_{12}$ & \ 2.07e$^{-01}$ &e$_{22}$& -7.17e$^{-02}$&e$_{32}$& -1.25e$^{-02}$ \\
			
			e$_{13}$&-4.91e$^{-01}$& e$_{23}$&-6.35e$^{-01}$ &e$_{33}$&\ 1.67e$^{-02}$ \\		
			e$_{14}$ & -3.86e$^{-01}$ &e$_{24}$& \ 3.05e$^{-01}$&e$_{34}$& \ 1.17e$^{+00}$\\
			e$_{15}$ & \ 1.49e$^{-01}$ &e$_{25}$& \ 1.45e$^{-01}$&e$_{35}$& \ 2.32e$^{+00}$ \\
			e$_{16}$& -4.74e$^{-01}$ &e$_{26}$& -1.14e$^{+00}$&e$_{36}$& \ 2.98e$^{-03}$\\
			\hline
		\end{tabular*}
	\end{table} 
	
	\par How successfully a piezoelectric material transforms electrical energy into mechanical energy or the other way around is measured by the electromechanical coupling coefficient (k$_{ij}$). It is computed by considering the piezoelectric tensors and other physical parameters. In order to determine the coupling constant values we utilize the following equation:
	
	\begin{equation}
		\textit{k}_{ij} = \dfrac{|\textit{e}_{ij}|}{\sqrt{C_{ij}\epsilon_{fs}\epsilon_0}}
		\label{Eq 7}
	\end{equation}
	
	\par where \textit{k$_{ij}$} being the electromechanical coupling coefficient, \textit{e$_{ij}$} represents the piezoelectric coefficients (in C/m$^2$). \textit{C$_{ij}$} is the elastic constants (in 10$^9$ Pa), with $\epsilon_{fs}$ as the permittivity of space and $\epsilon_0$ be the static dielectric constant at the zero stress level. Utilizing eqn \ref{Eq 7}, we calculate the electromechanical coupling constants of the highest piezoelectric responses from the surface slab systems and heterostructure under GGA and mGGA. 
	
	\begin{table}[hbt!]
		\small
		\caption{\ Calculated values of the piezo-electromechanical coupling constants for the surface slabs (001) under the employed functionals}
		\label{Table 6}
		\begin{tabular*}{0.48\textwidth}{@{\extracolsep{\fill}}l|l|l}
			\hline
			\textbf{Xc} &CsGeCl$_3$ (001)& RbSnBr$_3$ (001)\\
			\hline
			& \ \ \ \ \ \ e$_{ij}$, k$_{ij}$& \ \ \ \ \ \ e$_{ij}$, k$_{ij}$\\
			\hline
			\textbf{GGA}     &\ \ \ \ \  -1.36, 0.29 & \ \ \ \ \ -0.33, 0.27   \\
			\textbf{mGGA}     & \ \ \ \ \ -2.34, 0.85& \ \ \ \ \ -2.92, 0.99 \\
			\hline
		\end{tabular*}
	\end{table}
	
	\par Using the equation \ref{Eq 7}, we calculate the electromechanical coupling constants for GGA and mGGA based on the highest piezoelectric responses and other essential parameters. As listed in Table \ref{Table 6}, the obtained values of the coupling constants illustrate a high conversion rate for these materials.
	\section*{Conclusions}
	\par In this novel work, we have used the first principles approach to define and explore the significant attributes of the Pb-free halide perovskites in different phases. For all the phases studied, we employed the GGA and mGGA exchange functionals to treat the electron localizations. Herein, we observed an interesting result that all the investigated compounds happen to be semiconducting in nature. The surface slab cut (001) and constructed heterostructure are found to be energetically stable under the employed functionals. Our calculations show notably high and fair absorption coefficients $\sim$5.87-0.19
	$\times$ 10$^5$ cm$^{-1}$, from bulk to heterostructure, demonstrating that these materials are promising candidates for photovoltaic and optoelectronic devices. Due to their asymmetric orientations, interfacial electronic polarization emerges, leading to charge separation under an applied external force; giving rise to piezoelectric properties. Our calculated piezo-response from bulk$\rightarrow$surface$\rightarrow$heterophase exhibits a high response, reaching up to 2 C/m$^2$, indicating that the compounds under-investigated also are excellent candidates for piezoelectric applications.
	
	\section{Acknowledgments}
	\par L.Celestine acknowledges the Ministry of Tribal Affairs (Scholarship Division), Government of India for the support vide No.11019/07/2018-Sch with the Award No.202223-NFST-MIZ-01526.\\
	A. Laref acknowledges support from the "Research Center of the Female Scientific and Medical Colleges",  Deanship of Scientific Research, King Saud University.
\section*{Author Contributions}
\begin{itemize} 
\item L. Celestine: Formal analysis, Visualization, Validation, Literature review, Performed Calculation, Writing-original draft, writing-review \& editing.
\item R.Zosiamliana: Formal analysis, Visualization, Validation, writing-review \& editing.  
\item H. Laltlanmawii: Formal analysis, Visualization, Validation, writing-review \& editing.  
\item 	B. Chettri: Formal analysis, Visualization, Validation, writing-review \& editing.  
\item 	Lalhum Hima: Formal analysis, Visualization, Validation, writing-review \& editing.  
\item	Lalhriat Zuala: Formal analysis, Visualization, Validation, writing-review \& editing.  
\item 	S. Gurung: Formal analysis, Visualization, Validation, writing-review \& editing.  
\item A Laref: Formal analysis, Visualization, Validation, writing-review \& editing.  
\item D. P. Rai: Project management, Supervision, Resources, software, Formal analysis, Visualization, Validation, writing-review \& editing. 
\end{itemize}

\section*{Conflicts of interest}
	There are no conflicts to declare.
	
	\section*{Data availability}
	The data that support the findings of this study are available from the corresponding author upon reasonable request.
	
	\nocite{*}
	\bibliography{INTERFACE.bib}
	
\end{document}